\documentclass[conference]{IEEEtran}
\usepackage{makecell}

\usepackage{cite}
\usepackage{amsmath,amssymb,amsfonts}
\usepackage{algorithmic}
\usepackage{graphicx}
\usepackage{xcolor}
\usepackage{array}
\usepackage{tabularx}
\usepackage{comment}
\usepackage{amsthm}
\usepackage{url}
\usepackage{epstopdf}

\newcommand\blfootnote[1]{%
  \begingroup
  \renewcommand\thefootnote{}\footnote{#1}%
  \addtocounter{footnote}{-1}%
  \endgroup
}
\usepackage[normalem]{ulem}

\usepackage{booktabs}

\usepackage{tabularx}

\makeatletter
\def\ps@IEEEtitlepagestyle{%
  \def\@oddfoot{\mycopyrightnotice}%
}
\def\mycopyrightnotice{%
  \begin{minipage}{\textwidth}
  \centering \scriptsize
  \copyright 2022 IEEE. Personal use of this material is permitted. Permission from IEEE must be obtained for all other uses, in any current or future media, including reprinting/republishing this material for advertising or promotional purposes, creating new collective works, for resale or redistribution to servers or lists, or reuse of any copyrighted component of this work in other works.
  \end{minipage}
}
\makeatother

\begin{document}

\graphicspath{{./IEEE/figs/}}
\title{iTUAVs: Intermittently Tethered UAVs for Future 
Wireless Networks} 

\author{ \IEEEauthorblockN{Nesrine~Cherif\IEEEauthorrefmark{1}, Wael Jaafar\IEEEauthorrefmark{1}, Evgenii Vinogradov\IEEEauthorrefmark{1}, Halim Yanikomeroglu, Sofie Pollin and Abbas Yongacoglu}
}

\maketitle

\begin{abstract}
We propose the intermittently tethered unmanned aerial vehicle (iTUAV) as a tradeoff between the power availability of a tethered UAV (TUAV) and flexibility of an untethered UAV. An iTUAV can provide cellular connectivity while being temporarily tethered to the most adequate ground anchor. Also, it can flexibly detach from one anchor, travel, then attach to another one to maintain/improve the coverage quality for mobile users. Hence, we discuss here the existing UAV-based cellular networking technologies, followed by a detailed description of the iTUAV system, its components, and mode of operation. Subsequently, we present a comparative study of the existing and proposed systems highlighting the differences of key features such as mobility and energy. To emphasize the potential of iTUAV systems, we conduct a case study, evaluate the iTUAV performance, and compare it to benchmarks. Obtained results show that with only 10 anchors in the area, the iTUAV system can serve up to 90\% of the users covered by the untethered UAV swapping system. Moreover, results for a small case study prove that the iTUAV allows to balance performance/cost and can be implemented realistically. For instance, when user locations are clustered, with only 2 active iTUAVs and 4 anchors, achieved performance is superior to that of the system with 3 TUAVs, while when considering a single UAV on a 100 minutes event, a system with only 6 anchors outperforms the untethered UAV as it combines location flexibility with increased mission time.

\end{abstract}

\vspace{-5pt}
\section{Introduction}
\blfootnote{\hspace{-9pt}N. Cherif and A. Yongacoglu  are with uOttawa, Canada. W. Jaafar is with École de Technologie Supérieure, Canada, E. Vinagradov and S. Pollin are with KU Leuven, Belgium. H. Yanikomeorglu is with Carleton University, Canada. This work is funded by Huawei Canada, Natural Science and Engineering Research Council of Canada (NSERC), and Research Foundation Flanders (FWO), projects no. S003817N (OmniDrone) and G098020N. $^*$Authors have contributed equally.}
What are the advantages of UAV-mounted wireless communications? It increases the chances of a stable Line-of-Sight (LoS) dominated channel to ground terminals \cite{Alz2017}. UAVs can also move and optimize their locations to satisfy the dynamic service demand \cite{8247211,JaafarTVT}. These two features {}{represent} the core of all innovative wireless communication solutions {}{enabled by} Unmanned Aerial Vehicles (UAVs), including UAV base {}{stations (UAV-BSs), also called} on-board radio nodes (UxNB) {}{by} the 3$^{rd}$ Generation Partnership Project (3GPP).
The limited battery capacity is the main challenge to fully unleash the UAV-BS potential. To overcome this issue, several solutions have been proposed {}{such as} i) deploying recharging stations, 
ii) battery or UAV swapping, and iii) laser charging. Moreover, {}{providing UAVs} with permanent energy supply through a tether, a.k.a., tethered UAV (TUAVs), has been proposed as an alternative to the aforementioned solutions. Nevertheless, this concept offers unlimited flight time at the expense of limited mobility due to its predetermined hovering region \cite{9205314}.

\begin{figure}[t]
    \centering
    \includegraphics[trim={2cm 0 0 0},width=0.8\linewidth]{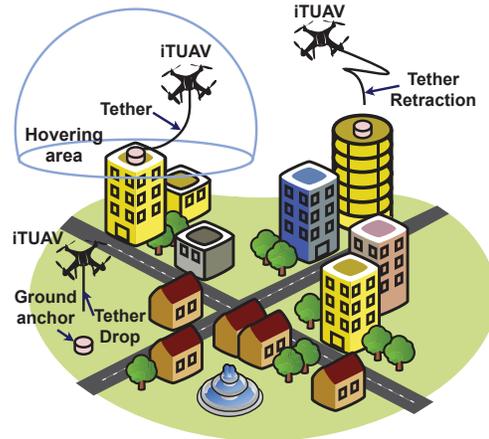}
    \caption{Proposed iTUAV system.}
    \label{fig:systemiTUAV}
\end{figure}

{}{Consequently,} we propose {}{here} a novel development to the operation mode of the TUAV system, which would make it more flexible and thus more attractive to industry. {}{It} consists on the introduction of an agile tether that can be attached/detached to/from existing infrastructures, called ground anchors, such as UAV and electric vehicle charging stations, illustrated in Fig.~\ref{fig:systemiTUAV}.
We call this novel system intermittently tethered UAV (iTUAV). It is noteworthy that ``ground anchor'' and ``anchor'' are used interchangeably throughout the paper.


\section{UAV-based Communications: A Review}
In recent years, the use of UAVs to provide cellular connectivity has attracted increasing attention and their potential to complement terrestrial networks has been extensively investigated. Due to their agility and flexibility, UAV-BSs temporary handle spikes in data demands during short-term events, such as large concerts and games \cite{irem}. 
However, UAV-BS operations are hampered by limited flying times, typically between 30 minutes and a few hours. To tackle this problem, numerous solutions have been investigated, e.g., battery recharging and swapping, UAV swapping, laser charging, and TUAV systems.

\subsection{Battery Recharging and Swapping}
To overcome the battery limitation, one can design a system where a UAV-BS flies back to a dedicated charging station to either recharge its battery or exchange it for a filled one \cite{Galkin2019}.

\emph{Critique:} Since UAV-BSs spend time to travel and recharge at a dockstation or land for battery swapping, these solutions cause frequent service interruptions, which is not ideal for stringent connectivity services, such as safety applications. 

\subsection{UAV-BSs Swapping}
In order to eliminate service interruptions, several works have proposed continuous UAV-BS swapping. In  \cite{ERDELJ2019101612} proposed automating {}{UAV swapping} for continuous cellular connectivity by monitoring {}{UAV-BSs} battery levels. 

\emph{Critique:} These approaches offer high UAV-BS deployment flexibility and sustained cellular services. However, {}{their} advantages come at the expense of higher capital and operational expenditures (CAPEX/OPEX), due to a larger UAV fleet.

\subsection{Laser Charging}
Another solution consists of recharging UAV-BSs on-the-fly through laser beaming. This technology can be implemented with a laser array oriented through a set of mirrors or diamonds aimed at the UAV-BS's collecting lens.
In spite of its great potential, laser beaming requires large collecting lenses and high laser power, two factors that limit {}{its} deployment, especially in urban areas. Alternatively, distributed laser charging (DLC) involves the use of photo-voltaic cells instead of a collecting lens, which is practical and cost-effective for small UAV-BSs \cite{jaafar2021}. Unlike laser beaming, DLC uses less than {}{one} kilowatt transmit power, {}{thus favoring its deployment} anywhere. Finally, efficient DLC requires LoS between the charging source and UAV-BS.

\emph{Critique:} Due to health-related concerns, laser beaming with high power cannot be used in densely populated areas. {}{In the meanwhile, DLC may} perform poorly in non-LoS conditions. Moreover, due to its slow charging regime, DLC is only efficient when the UAV-BS is on stand-by \cite{jaafar2021}.

\subsection{TUAV Systems}
TUAV {}{does not suffer from} short operation time. {}{Its} main idea involves linking the UAV to the ground using a tether that provides both power supply and {}{high-capacity backhaul}. Consequently, the limited TUAV mobility is compensated by a longer operation time and higher backhauling capacity. 
\cite{9205314} and \cite{9091128} introduced specific constraints to TUAVs such as limited tether length and safety considerations to avoid tangling the tether {}{around} surrounding buildings.
Also, they assumed that the tether serves both as power cable and wired backhaul link. This assumption implies complex and high cost tether and  anchor designs. Alternatively, a simpler design can support power supply only, while relying on wireless backhauling through either radio or free-space optics \cite{8999435}.

\emph{Critique:} The main disadvantage is the reduced flexibility due to highly restricted flying distance. Moreover, backhauling through a tether implies higher equipment cost.
Thus, they may be considered a good alternative to terrestrial BSs (TBSs) and by far do not offer the same advantages as UAV-BSs.

\section{Intermittently Tethered UAVs}
Given the limited hovering areas of TUAVs and battery-limited operation of untethered UAVs, we propose here a compromise solution, a.k.a. iTUAV, which offers both the flexibility of a UAV and the sustainability of a TUAV. In what follows, we present the iTUAV system components, operation mode, and a qualitative features comparison to existing technologies.

\subsection{System Components}
An iTUAV-aided communication system involves several components shown in Fig. \ref{fig:systemiTUAV} and described as follows: 

\paragraph{iTUAV} An iTUAV is a small UAV equipped with a lightweight 5G New Radio (NR) UxNB for communication, {}{e.g., Ericsson's network-on-a-drone}. Specifically, the iTUAV has the following {}{modules}: 
    \begin{itemize}
    \item \textbf{Access:} 
    {}{As specified in 3GPP TR 22.829, UxNB is able to connect to the 5G core network and operate as a BS, via a wireless backhaul link, and {}{can }bootstrap as a BS, from the core network's perspective. To do so, we assume that the UxNB is equipped with multiple antennas enabling directional patterns towards users \cite{9793669}.}
    \item \textbf{Backhaul:} The backhaul can be provided by a wireless gateway (e.g., a nearby TBS, TUAV, or satellite) or through an associated anchor, if it provides backhauling services. In the case of a wireless backhaul, integrated access and backhaul (IAB) can be leveraged \cite{8999435,9793669}. Since 3GPP does not enforce any particular IAB implementation, radio access and backhauling can use different frequency bands or the same spectrum. The latter is possible since a {UxNB} supports directional antenna arrays {to spatially separate} the links. 

        \item \textbf{Power:} To reduce costs, we assume that the iTUAV is equipped with a \textit{small battery} that enables short flight times to reach  anchors and {}{re}connect through a tether. Moreover, each iTUAV is equipped with a \textit{socket} that allows it to attach a retractable tether from the anchor and that can be used simultaneously to power the motors in-flight and to recharge the battery. 
        This design suggests {}{landing} over the anchor to attach {}{the iTUAV}, {}{thus causing} service delays. Alternatively, the iTUAV can be equipped with a \textit{lightweight tether} that is dropped vertically to the anchor rather than landing on it \cite{9469424}.

    \end{itemize}

\paragraph{Ground anchors} These are mainly power supply sources used to sustain iTUAV operations. A ground anchor may include one or several of the following components. First, it must have access to the power grid and be equipped with adequate \textit{power converters} to support different types and sizes of UAVs and electric vehicles. Second, it may be provided with \textit{standardized sockets} to plug UAVs and electric vehicles. Third, it may include \textit{power tethers} dedicated for iTUAVs with power sockets. Fourth, it may have a \textit{battery swapping station}, which may be used by UAV-BSs. Additionally, the  anchor can offer backhauling  through a \textit{wired link} combined with \textit{on-site routing equipment}.  
The versatility of the anchor role favors its business and practical interests as part of the smart city infrastructure. For instance, anchors can be easily and seamlessly integrated to lampposts, road signs, and lights.

\paragraph{Gateway} This is the entry-point of the iTUAV to the core network. It can be the routing equipment of an anchor, a TBS, a TUAV, or a satellite. Except for the anchor that allows a direct wired connection to the core network, the other alternatives require wireless links with the iTUAV. Although wireless transmissions are less efficient than wired ones, {}{communicating} with a TBS equipped with \textit{highly directional antenna arrays} {}{using} IAB is an effective and {}{cheap}
alternative.

\begin{figure*}[t]
    \centering
    \leavevmode
    \includegraphics[width=0.9\linewidth,trim=20 328 155 421, clip]{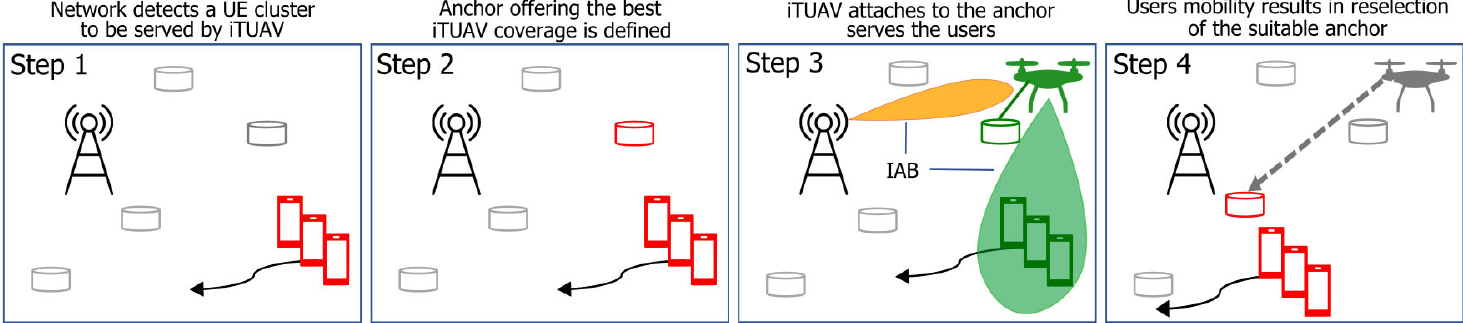}
    \caption{{Operation steps of the iTUAV system (the solid black arrows show the users' motion direction; the gray dashed arrow represents the ground anchor selection; the green solid line is the tether that connects the iTUAV to the anchor; the green and orange beams are the access and backhaul links, respectively. }}
    \label{fig:operation}
\end{figure*}

\subsection{System Operation}
{}{We assume a communication system deployed in an urban environment, e.g., smart city, consisting of iTUAVs, TBSs, ground anchors, and mobile ground users. The main goal of an iTUAV-aided communication network is to meet the performance requirements related to ubiquitous coverage, e.g., during temporary events, and to on-demand network densification due, for instance, to peak traffic demands.} 
Fig. \ref{fig:operation} details the operational steps of {}{this} system, explained below:

\begin{enumerate}
    \item A TBS detects users that cannot be reliably served. These users are clustered then assigned to an iTUAV.    
    \item To serve users' clusters, the optimal location of the iTUAV and its associated  anchor are {}{jointly} determined.
    The formulation and solution of this problem may take into account limitations, such as the propagation environment, backhaul link quality and TBS resources. 
    
    \item The iTUAV flies to the anchor and connects to it through a retractable power tether. Then, it travels to its optimal location and hovers to serve the associated cluster.
    \item Since users are mobile, the optimal location of the iTUAV (determined in the previous steps) may become out of reach of the tether's length. In this case, the new optimal location and associated anchor must be recalculated. Subsequently, the iTUAV detaches itself from the current anchor and {}{re-executes} step 3.
\end{enumerate}

\subsection{iTUAV Systems Versus Existing Technologies}
The idea behind iTUAV emerges from the tradeoff between {}{mobility and power availability}. In what follows, we compare the UAV-based cellular networks, namely iTUAV, UAV (i.e., untethered UAV-BS), TUAV, and TBS:

\begin{itemize}
    \item \textbf{Mobility:} UAVs equipped with BSs gained popularity thanks to their exceptional {}{3D} mobility. A typical UAV-BS use-case consists of providing connectivity for temporary events with high traffic demands. Such scenarios cannot be handled by terrestrial cellular networks due to the fixed nature of TBSs. In contrast, TUAVs and iTUAVs are much less constrained, and thus {}{can} leverage their partially limited mobility to better serve users.
    Nevertheless, a TUAV is permanently fixed to an anchor, which limits its region of activity, compared to an iTUAV, which is able to travel from one anchor to another. Consequently, iTUAVs cover wider areas than TUAVs.
    \item \textbf{Re-deployment Flexibility:} UAVs are valued for their {}{rapid deployment feature, due to their high flying speed}. 
    An iTUAV can be re-deployed almost as quickly, either by relocating within a hovering region, subject to the tether length, or by traveling to the closest anchor to the {}{targeted} location.
    {}{Note} that the inherent re-location flexibility of a TUAV is strongly linked to the length of the tether and thus to its hovering region. That means, if redeployment is required in an area outside the hovering region, the TUAV cannot satisfy the new requirement. Finally, TBSs lack this feature due to their fixed nature.

    \item \textbf{Cellular Reliability:} 
    A terrestrial network is considered reliable when its {}{TBSs} deployment is optimized and does not {}{experience} substantial coverage gaps. An aerial BS increases the probability of LoS {}{to communicate} with ground users and thus is capable of providing a similar performance as a TBS. However, {}{untethered} UAV-BSs rely on wireless backhauling, which is limited compared to a TBS wired backhaul. A TUAV system offers better reliability due to its wired backhaul; {}{nevertheless}, its limited motion prevents it from offering a highly reliable coverage. Finally, an iTUAV ensures a better cellular reliability than {}{UAV/TUAV} systems due to its flexibility in connecting with different anchors to achieve better {}{access/backhaul} links. With enough available anchors, iTUAVs can achieve similar performance to TBSs.

    \item \textbf{Backhaul Availability:} A terrestrial network provides a highly available backhaul through fiber links and point-to-point RF transmissions. Similarly, {}{TUAV-based} solutions assume that {}{backhauling is} carried through fiber links aggregated with the power supply within the tether. 
    By contrast, iTUAVs are expected to operate opportunistically with both wireless and wired {}{backhauling}. Indeed, whenever the anchor provides a tether with a {fiber optic} data link, backhauling {}{goes} through it. However, if an anchor is exclusively a power source or when an iTUAV is traveling between anchors, then it has to rely on wireless links to {most adequate} TBSs or {}{to} other wireless gateways. Finally, {}{untethered} UAV-BSs rely solely on {}{the} wireless backhaul.

    \item \textbf{Energy Availability:} Clearly, the untethered UAV-BS has the lowest {}{power capacity}. By contrast, energy for TBSs and TUAVs is available continuously. iTUAVs, although occasionally traveling, enjoy a very high energy availability. When an iTUAV is attached to an anchor, it has unlimited access to power {}{as} TUAVs. This energy is used for both iTUAV operation and battery recharging. However, when it is traveling between anchors, it relies exclusively on its on-board battery. Note that {}{locations} of anchors and {}{battery} capacity must be carefully and jointly designed in order to avoid draining the iTUAV's battery when traveling between anchors.

    \item \textbf{Infrastructure Reuse:} Terrestrial cellular infrastructure suffers from a low usage rate due to users {}{with} sporadic traffic demands. In TUAV systems, the anchor is permanently powering a single or several TUAVs through tether(s), resulting in small or no opportunities to reuse the tether or the power supply of the  anchor. By contrast, an iTUAV is only attached to an anchor for the time of its mission, then it releases the tether and flies to a different location. The low usage rate of anchors by iTUAVs favors their reuse for other applications such as recharging flying taxis, cargo UAVs, and electric vehicles. Moreover, the fleet of iTUAVs is much smaller than the one considered by UAV {}{swapping}. 
    Infrastructure reuse substantially increases {}{iTUAVs'} economic viability. Moreover, the additional generated revenue from {}{anchors reuse} can offset the investment of extensive anchor networks and the expenses thereafter (e.g., rental of rooftops).
    
    \item \textbf{CAPEX:} 
    One of the main concerns in setting up cellular services is the amount of initial investment needed.
    {}{TBSs} are usually costly. Locations for their deployment need to be purchased or leased, and they require expensive electrical and communications {}{equipment}.
    In contrast, {}{UAVs} are less expensive, but {}{have limitations that need to be considered.}
    This typically leads to {}{purchasing} a high number of UAVs, batteries, and recharging {}{stations} to enable UAV swapping, battery {}{recharging/swapping}. In the case of TUAVs, the initial investment consists of {}{buying} UAVs with their tether anchors for power supply and communication. Due to the limited tether length, covering a large area would require the deployment of several {}{TUAVs,} which may rapidly increase costs. Like TUAVs, iTUAVs rely on anchors, 
    but with lighter functionalities. Indeed, 
    Compared to TUAV systems, a smaller number of iTUAVs but a slightly higher number of anchors can sustain wireless services in large areas.  
    
    \item \textbf{OPEX:} 
    Terrestrial networks still face high operating costs {}{for} equipment maintenance. The use of UAVs instead requires sophisticated software and professional UAV pilots to lead missions and organize {}{recharging/swapping} operations. For {TUAV and iTUAV systems}, {in addition to operation software,} maintenance of UAVs and anchors is needed. {However, the cost of maintaining a UAV is expected to be higher than that for an anchor. Hence,} OPEX might be slightly lower for {}{iTUAVs} due to {}{their} forecast smaller number, but slightly higher number of anchors. Finally, iTUAV anchors can generate additional revenue through their {}{reuse}, which is expected to significantly offset {}{the} CAPEX/OPEX costs.      

\end{itemize}

Based on the foregoing {}{discussion}, iTUAV systems present great potential to compete with or complement existing cellular technologies. They represent a promising middle-solution between the mobility features of UAVs and the operational sustainability of TUAVs and {}{TBSs}.

\section{Case Study}
To corroborate our findings on iTUAV systems, we investigate here their achievable performance in terms of average number of covered users as a function of terrestrial users density and of its sustainability. These results are compared against those of UAV and TUAV systems.

\subsection{Simulation setup}
We consider a $3\times 3$ km$^2$ urban area where a  {}{single }UAV (or a TUAV or an iTUAV) is deployed to provide cellular coverage for terrestrial users equipment (UEs). The transmit power is set to $P_t=30$ dBm and the UE receiver sensitivity is $P_{\min}=-70$ dBm \cite{Alz2017}, which translates to a path loss threshold of $100$ dB. We assume that the air-to-ground channels follow the probabilistic line-of-sight (LoS) and non-LoS (NLoS) propagation model \cite{6863654}, operating at frequency $f_c=2$ GHz. The average height of anchors is $50$ m and the maximal tether length for both the TUAV and iTUAV is set to $150$ m \cite{9091128}. For the sake of simplicity, we assume that the TUAV (resp. iTUAV) minimum inclination angle that prevents tangling around surrounding buildings and ensures the safety of the tether is set to 0$^\circ$. The UEs are distributed according to the Thomas point process. With this distribution,  clustering of UEs is  modeled  through the  coefficient of variation (CoV) of  the Voronoi cell area around each user \cite{Mirahsan2015}. For instance, when $\text{CoV}=1$, the UEs are uniformly distributed according to the Poisson point process, whereas a higher CoV represents more clustered users around a virtual cluster head. At each realization, i.e., {}{UEs and anchors geographical distributions}, the iTUAV determines the optimal anchor and attaches to it to provide cellular coverage to the highest number of UEs. In contrast, the untethered UAV freely determines and reaches the optimal 3D location that maximizes its coverage \cite{Alz2017}, whereas the TUAV, constrained to its pseudo-static deployment, relocates withing its hovering region to cover the maximal number of UEs \cite{9091128}. Finally, results are averaged over both the UEs' geographical distribution and anchor positions in each iteration.

Given the identified constraints of iTUAV (resp. UAV and TUAV) system, the formulated problems aim to maximize the cellular coverage of iTUAVs (resp. UAVs and TUAVs) through the optimization of the {}{UAV-BS} 3D location and ground anchor association (iTUAV). Although seeming different, these problems are convex \cite{Alz2017,nesrine2020globecom}. Hence, they can be solved using MOSEK parser in the CVX package of Matlab.

\subsection{Results and Discussion}

\begin{figure*}[t]
     \begin{minipage}{0.3\linewidth}
     \includegraphics[trim={0cm 0cm 0cm 0.8cm},width=165pt]{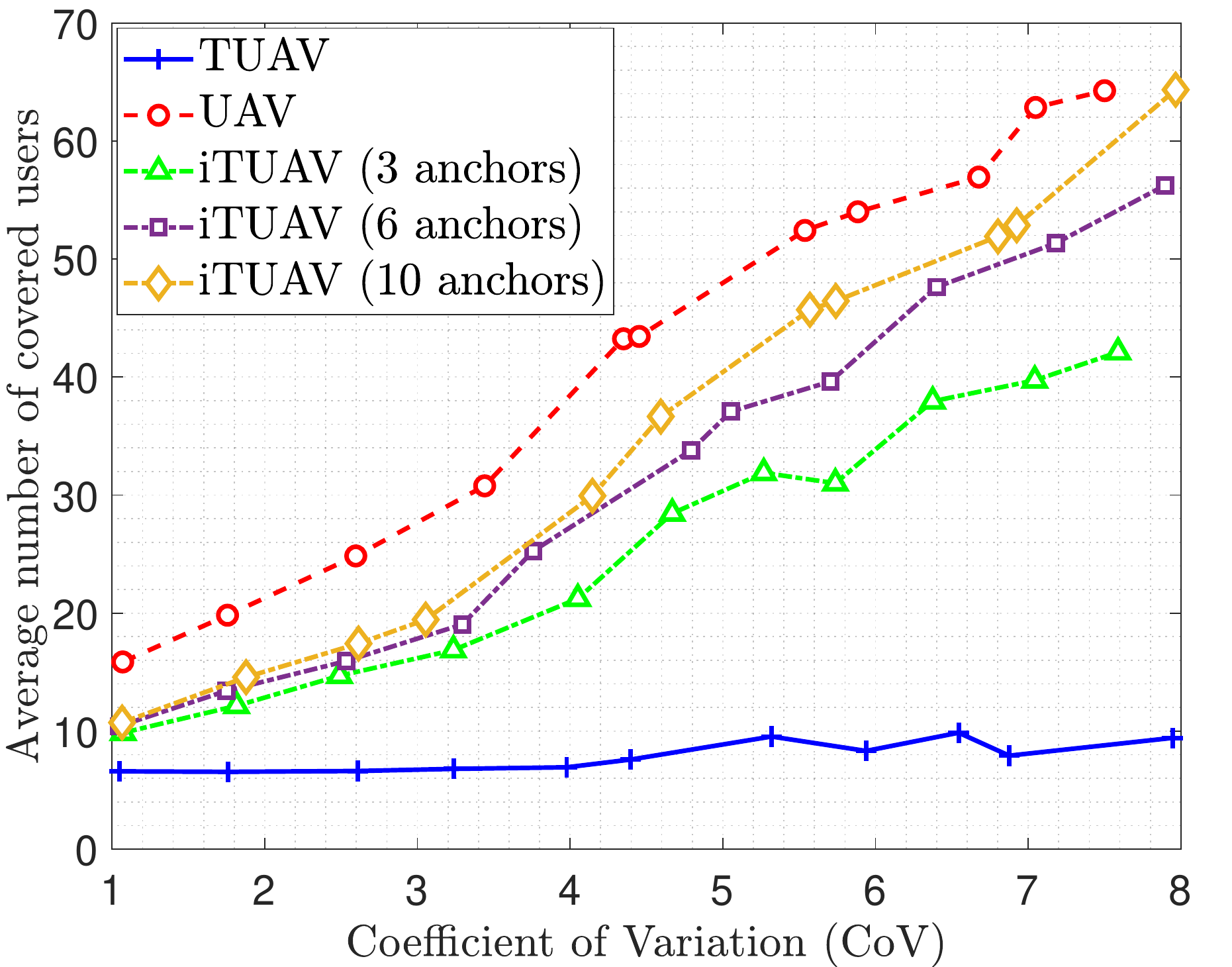}
       \caption{Coverage performance of UAV {(with swapping)}, TUAV, and iTUAV systems, for different user distributions.}
    \label{fig:covered_users}
     \end{minipage}
     \hfill
     \begin{minipage}{0.3\linewidth}
   \includegraphics[trim={0cm 1cm 0cm 0cm},width=165pt]{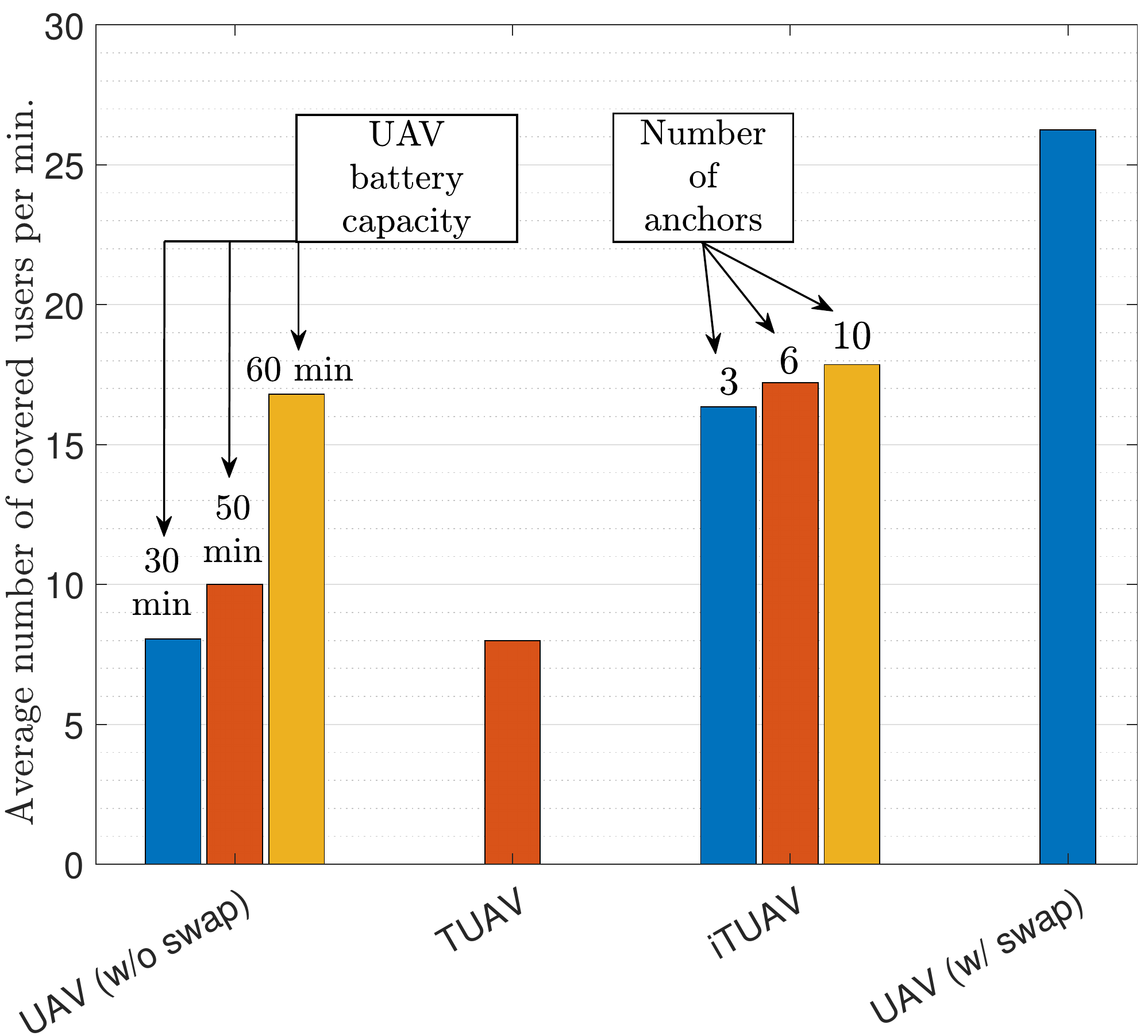}
	\caption{Coverage performance of UAV, TUAV, and iTUAV systems for a temporary event of 100 minutes ($\text{CoV}=3$).}
	\label{fig:service}
     \end{minipage}
     \hfill
     \begin{minipage}{0.3\linewidth}
    \includegraphics[trim={0.2cm 1cm 0cm 1.4cm},width=165pt]{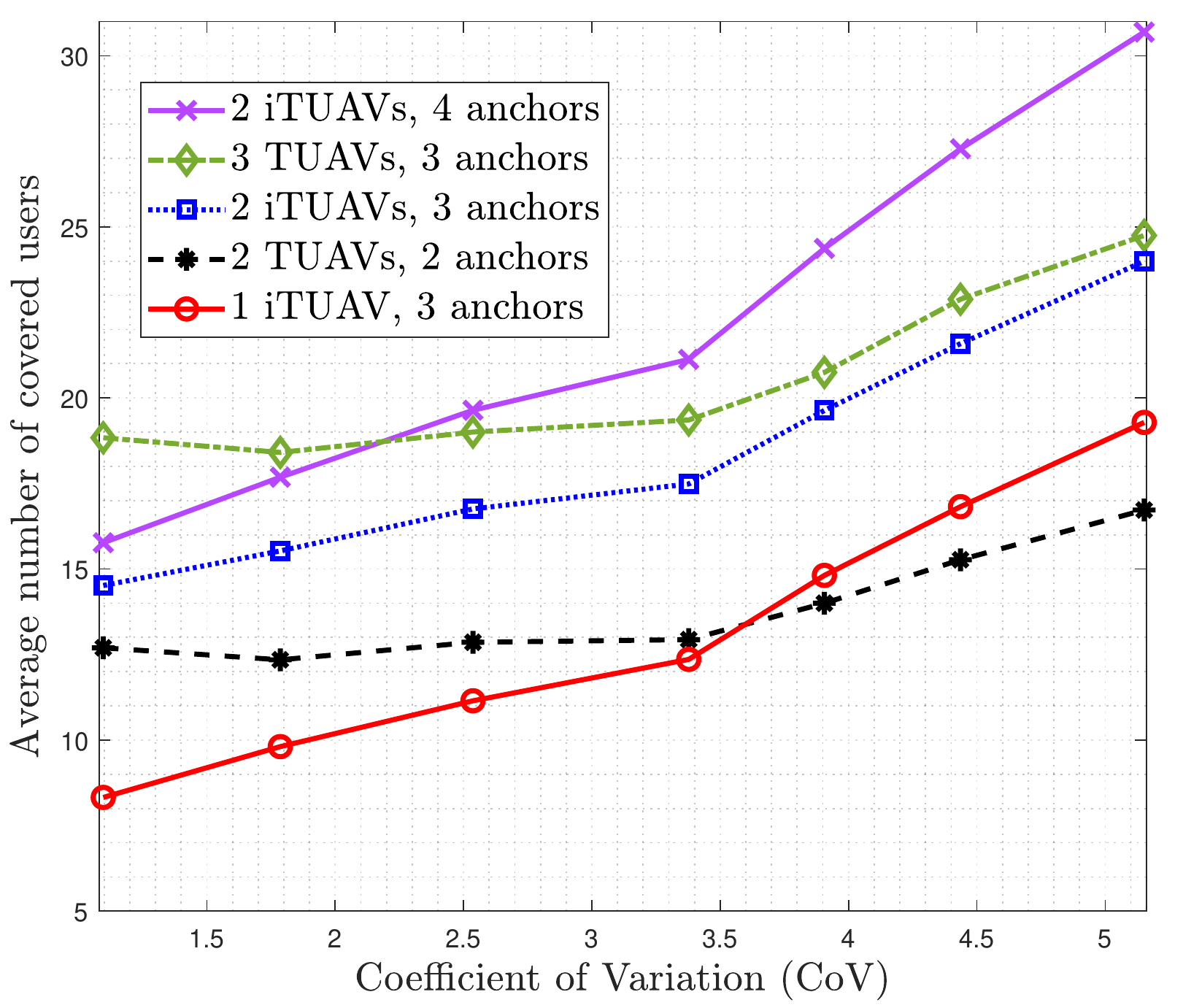}
	\caption{Coverage performance of TUAV and iTUAV systems (different setups).}
	\label{fig:TUAV}
	\end{minipage}
   \end{figure*}

Fig. \ref{fig:covered_users} shows the average number of users {}{covered by the single serving UAV} as a function of the CoV for different UAV systems, namely, UAV, TUAV, and iTUAV. {Although TUAV and iTUAV deploy a single UAV, the UAV system assumes UAV swapping in order to ensure a similar service continuity over time \cite{ERDELJ2019101612}}. As expected, the UAV presents the best coverage performance {}{thanks to} its optimal placement. By contrast, the TUAV is extremely limited in its flying range by the tether/anchor location, thus often missing the best placement achieved by the UAV. Alternatively, the iTUAV outperforms the TUAV due to the availability of several anchors, which allows it to select the best one and optimally place itself to cover a higher number of users. Nevertheless, the iTUAV system performs below the UAV, since it may miss the optimal placement if none of the anchors is located close enough (subject to the tether's length) to the UAV optimal location. With more anchors, the iTUAV exploits near-optimal hovering locations, and thus achieves better coverage performance. Finally, as the CoV increases, coverage improves for all systems. Since UEs are more clustered, UAV communicates through better channels to a higher number of them.

Fig. \ref{fig:service} compares the performances of the UAV {}{without swapping (w/o swap), UAV with swapping (w/ swap)}, TUAV, and iTUAV systems in terms of the average number of covered users per time unit (min.). {Specifically, we define a cellular radio access mission consisting of deploying {}{one UAV} within a target area to provide cellular connectivity during a {}{temporary event} of 100 minutes.\footnote{{}{This assumption implies no UAV swapping for the untethered UAV and the mission of the UAV is hence limited by the battery lifetime}.}} During this event, we aim to maximize the average number of connected users to the mobile UAV. The average is calculated over multiple network topologies (i.e., different scenarios for users' and anchors' locations) and time.
We assume different battery capacities for the UAV {}{(w/o swap)}, ranging between 30 and 60 minutes of flying time, while the iTUAV relies on a number of anchors, between 3 and 10. We notice that the UAV {}{(w/o swap)} coverage performance is low with a small battery capacity (e.g., 8 users per min. with a battery of 30 min.), but it improves with a larger battery capacity (e.g., 16.8 users per min. with a battery of 60 min.).
However, the TUAV can continuously serve 8 users per minute. The TUAV performance is mainly limited by the tether that prevents the UAV from hovering at a better location where it could serve more users. With more anchors, the iTUAV enjoys better performance over time as it can also serve more users owing to its mobility and stable power source, i.e., through the tether and the on-board battery when traveling between anchors. {}{Finally, considering UAV (w/ swap), this system is equivalent to having an unlimited battery capacity while enjoying full motion flexibility, thus it achieves the best performance, i.e., around 26 users in average are covered through time. This performance, however, comes at the expense of an additional investment in a higher number of UAVs and more complex operations.} In a nutshell, we conclude that, thanks to its unlimited power and flexible association to anchors, the iTUAV is able to sustain a reliable cellular service over time, {}{which is} better than the UAV {}{(w/o swap)} and TUAV systems, {}{but slightly below that of the UAV (w/ swap) reference system. If a higher number and more adequate locations of anchors are available, iTUAV performance is expected to reach that of the UAV (w/ swap)}.

In Fig. \ref{fig:TUAV}, we illustrate the cellular coverage performance of iTUAV and TUAV systems with multiple UAVs as function of the CoV. When the users are very clustered, the iTUAV system performance mainly depends on the number of anchors, and a system with only 1 active iTUAV but 3 anchor locations even outperforms a system with 2 fixed TUAVs. A system with only 2 iTUAVs and 4 anchor locations, clearly achieves the best performance in clustered users scenarios. This is due to the opportunistic iTUAV behavior to move freely between anchors and associate with the best one. In contrast, the TUAV system has less leverage on relocating due to its static deployment.

\section{Challenges and Future Directions}
As  research of TUAV/iTUAV systems is in its infancy, several challenges remain. Next, we discuss some key issues:

\begin{itemize}
    \item \textit{New anchor, UAV, and tether designs:} Current TUAVs are built with the tether already attached at both anchor and UAV ends, such that attaching/detaching the tether must be executed manually. To achieve more flexibility to iTUAVs, two options can be considered: 1) the tether is within the  anchor and it provides an easy-clip attach/detach system, but would require landing over the anchor to connect, or 2) the tether is within the iTUAV. The iTUAV drops the tether and the anchor ``catches'' it to establish the connection. The first option would require a longer power connection time than the second due to additional landing and taking-off, while the second option would require a careful ``cable catching'' mechanism to bypass environmental factors, such as strong winds and obstacles. The second option would also require a lightweight and inexpensive tether design to reduce impact on the UAV's size, weight, power and cost. For this matter, industry players, such as VICOR, are proposing to transfer high voltage over thin and lightweight tethers and to use small fixed-ratio bus converters within UAVs. Moreover, depending on whether the iTUAV will be equipped with a very lightweight tether or not, the challenge is to design automatic hold/release mechanical or electromagnetic winches and sockets at the  stations and/or UAVs. Currently, similar conductive automatic charging systems are being prototyped for electric vehicles, such as the DAZEplug, which can inspire equivalent systems for iTUAVs. In addition, EasyAerial is on the right track with its Raptor UAV, which is capable of detaching itself from the tether. 
    
    \item \textit{Management of 
    iTUAV operations:}
    The management of UAV operations is an active area of research a in the context of unmanned aircraft system traffic management (UTM). For iTUAVs, operations are slightly different since they can ``jump'' from one anchor to another and thus induce unpredictable environmental changes.
    This aspect should be considered when integrating iTUAV operations within UTM systems. Also, iTUAV operations may be affected by the deployment strategy of anchors. For instance, vehicle-mounted anchors may change location, thus causing iTUAV placement/trajectory updates. 
    
    \item \textit{Integration of iTUAV with 5G and beyond technologies:}
    Although the proposed iTUAV system offers a tradeoff between the flexibility of UAVs and the unlimited power of TUAVs, their performance is still suboptimal. To further enhance iTUAV communications, other technologies can be leveraged, such as reconfigurable smart surfaces (RISs) and device-to-device (D2D) communications. Specifically, RIS and D2D can extend the coverage area of an iTUAV and thus reduce the number of required  anchors. Such approaches are expected to drastically reduce the deployment costs of iTUAV systems.

\end{itemize}

\section{Conclusion}
With the proliferation of UAV-based applications and the recent development of TUAVs, iTUAV systems are envisioned in this article as an intermediate solution that offers a tradeoff between the continuous power supply of TUAVs and the mobility of UAVs. In this paper, we first provided an overview of UAV-based communications, including UAV-BSs and TUAVs. Then, we proposed the iTUAV system and detailed its components, operation mode, and distinguishing features. Through a case study, we evaluated the coverage performance of the proposed iTUAV system and compared it to UAV-BS and TUAV counterparts. Preliminary results demonstrated the potential of iTUAV systems to sustain long and reliable cellular connectivity missions. However, several issues need to be addressed before their extensive deployment.
     

\bibliographystyle{IEEEtran}  
\bibliography{references}







\end{document}